\documentstyle[prl,aps,amsfonts,amssymb,epsfig]{revtex}
\begin{document}
\draft 
\twocolumn[\hsize\textwidth\columnwidth\hsize\csname 
@twocolumnfalse\endcsname
\title{Inversion Symmetry in the Spin-Peierls Compound $\alpha^{\prime}$-NaV$_2$O$_5$}
\author{A. Meetsma, J.L. de Boer, A. Damascelli, and T.T.M. Palstra}
\address{Materials Science Center,  University of Groningen, Nijenborgh 4,
 NL-9747 AG Groningen, The Netherlands.}
\author{J. Jegoudez, and A. Revcolevschi}
\address{Laboratoire de Chimie des Solides, Universit$\acute{e}$ de 
Paris-sud, B$\hat{a}$timent 414, F-91405 Orsay, France}
\date{January 23, 1998}
\maketitle
\begin{abstract}
At room-temperature $\alpha^{\prime}$-NaV$_2$O$_5$  was found to have the centrosymmetric
 space group {\em Pmmn}. This space group implies the  presence of only one kind of
 V site in  contrast with previous reports\cite{carpy} of the 
 non-centrosymmetric counterpart {\em P2$_1$mn}. 
 This indicates a non-integer valence state of vanadium.
 Furthermore, this symmetry has consequences for the interpretation of the 
 transition at 34 K, which was ascribed to a spin-Peierls transition of one 
 dimensional chains of V$^{4+}$.
\end{abstract}
\vskip2pc]
\narrowtext

\section {Comment}

Low dimensional quantum systems have revealed in recent years many new
 properties in their magnetic and electronic transport behavior.  The 
 vanadates complement in many respects the copper-oxide systems, with 
 spin-less, S=0, 1/2 and 1 states, obtained by V$^{5+}$ (d$^{0}$), V$^{4+}$ (d$^{1}$), 
 and V$^{3+}$ (d$^{2}$), or Cu$^{1+}$ (d$^{10}$), Cu$^{2+}$ (d$^{9}$), and Cu$^{3+}$ (d$^{8}$), 
 respectively.  
 Recently, the first observation 
 of a spin-Peierls (SP) transition in an inorganic compound in CuGeO$_3$\cite{hase} 
 was complemented by the observation of a  SP 
 transition in $\alpha^{\prime}$-NaV$_2$O$_5$\cite{isobe}. 
 Evidence for this one-dimensional magnetic behavior was found in the 
 temperature dependence of the susceptibility, which can be accurately 
 described at high temperatures by the Bonner-Fisher\cite{fisher} model for a S=1/2
 linear chain. Below T$_{SP}$=34 K an isotropic exponential decrease in the
 magnetic susceptibility is observed, as evidence for a non-magnetic ground
 state. The magnetic behavior of the SP state is complemented by SP
 signatures in a number of other measurements.

The usual interpretation of the transition at 34 K as SP transition
 is based on the structure determination of Carpy and Galy\cite{carpy}. Their 
 refinement, based on photographic data, in the non-centrosymmetric space group
 {\em P2$_1$mn}, allows for two vanadium positions in the asymmetric unit. 
 These sites were interpreted with different valence states, V$^{4+}$ and V$^{5+}$. 
 In their model the resulting one dimensional S=1/2  V$^{4+}$ chain can cause
 the observed Bonner-Fisher-like temperature dependence in the magnetic
 susceptibility, and a SP transition at low temperature. 
 However, the reported structure determination of Carpy and Galy yielded atomic
 coordinates with a pseudo inversion center at (0.259, 0.25, 0.11)\cite{lapage}. 
 Therefore, we have undertaken a structure  redetermination to investigate the (centro)symmetry. 

The structure can be constructed from double-rows of edge sharing pyramids, 
 one facing up and the other down. These double rows are connected by sharing 
 corners, yielding a planar material. These planes are stacked with the Na
 in the channels of the pyramids, as shown in Fig.~1. The eight-fold 
 coordination of Na is somewhat more symmetric with Na-O distances 
 (2.4325(11)-2.6038(9) \AA) than in the refinement of Carpy and Galy
 (2.43-2.90 \AA)\cite{carpy}.

 The $\alpha^{\prime}$-NaV$_2$O$_5$ structure it thus similar to that of 
 CaV$_2$O$_5$\cite{onada}. In CaV$_2$O$_5$ the V-O bond distances in the base of the
 square pyramids have a smaller range (1.90-1.98 \AA\ {\em versus} 1.8259(6)-1.9867(9) \AA). 
 However, in this material the valence state of the V is
 uniformly 4+, making this material an interesting spin ladder-like compound. 
 
 We think that our structure determination is proof for the space group {\em Pmmn}.
 Our evidence is the very low R$_F$-value of 0.015, and the fact that no lower
 value of R$_F$ can be obtained when omitting the inversion center. 
 The noncentrosymmetric space group {\em P2$_1$mn} reported by Carpy and Galy\cite{carpy} is,
 in our opinion the result of the limited data set of 117 reflections, and the
 photographic-data quality. 

 Nevertheless it is worthwhile to assess the validity of very small distortions
 yielding lower symmetry. The  centrosymmetric space group {\em Pmmn} indicates a 
 non-integer valence of V, which contrasts with the observation of a 
 SP transition and with the observed optical band gap of $\sim$1 eV. 
 Therefore, we considered refinement in the polar 'equivalent' of {\em Pmmn}, {\em i.e.}, 
 the non centrosymmetric space-group {\em P2$_1$mn}, in greater detail. 
 This analysis shows that the standard deviations of the atomic positions in
 the polar refinement are approximately a factor ten larger than in the 
 centrosymmetric refinement. This indicates much shallower minima in the 
 least-squares refinement, caused by large correlation between atomic 
 coordinates related by the pseudo-inversion-center. Similarly, the 
 least-squares refinement protocol yields a substantial number (15) of 
 large ($>$ 0.90) correlation-coefficients between various parameters. 
 Most equivalent bonds in {\em Pmmn} are in {\em P2$_1$mn} still almost equal. 
 The largest difference in interatomic distances between formerly equivalent
 bonds is found for V-O1. The bond length of 1.8259(6) \AA\ in {\em Pmmn} splits 
 into 1.7966(63) \AA\ and 1.8543(63) \AA\ in {\em P2$_1$mn}, {\em i.e.}, a
 displacement of 0.0289 \AA\ from the average value of 1.8254. 
 One can interpret this measure of noncentrosymmetry of 0.029 \AA\ in two ways. 

 The common 'crystallographic' interpretation considers the noncentrosymmetry 
 as an artifact. Table 2 shows that V and O's have all anisotropic 
 temperature ellipsoids. It is well known that one can mimic this by making 
 the space group noncentrosymmetric and using a more isotropic temperature 
 factor. Obviously, this causes large correlation between parameters in the
 least-squares procedure. Furthermore, one should keep in mind that the
 calculated standard deviations are based on random fluctuations, and are
 significantly underestimated when correlation is important. 
 Therefore, this interpretation assigns a much higher probability to the 
 centric space group {\em Pmmn}. 

 An alternative interpretation is to use the statistics not to distinguish
 between the symmetries, but to quantify the maximum deviations from 
 centrosymmetry in order to asses, {\em e.g.}, the magnitude of the transition 
 dipole moment. Our analysis shows that the reported standard deviations for
 the atomic positions are based on the underlying symmetry. This allows
 0.029 \AA\ deviations from centrosymmetry, about five times larger than the 
 calculated e.s.d. in the interatomic bond length, and should therefore be
 considered significant. 

 We conclude that our data are evidence for the centrosymmetric space group
 {\em Pmmn}. Deviations from centrosymmetry are unlikely but cannot be excluded up 
 to 0.03 \AA. The Flack x parameter is often used to indicate
 noncentrosymmetric structures. However, its value of 0.41(7) indicates at
 most twinning in case of a noncentrosymmetric structure, and this would still 
 lead to a centrosymmetric 'space-average'. 

 While we have investigated the crystallographic structure of $\alpha^{\prime}$-NaV$_2$O$_5$,
 its electronic structure is not so obvious. Clearly, the original
 assignment of different valence states of V$^{4+}$ and V$^{5+}$ needs modification.
 Furthermore, the interpretation of the transition at 34 K as a SP
 transition requires a different model. Still, the 1-dimensional behavior of 
 the magnetic  susceptibility and the insulating properties of this non-integer
 valent material need to be incorporated in this theory.
 Also, the higher symmetry that we propose should be consistent with the 
 symmetry as observed in, {\em e.g.}, Raman spectroscopy and IR spectroscopy.  
 Further study of the electronic properties of $\alpha^{\prime}$-NaV$_2$O$_5$ is in progress.

\section {Experimental}

Crystal growth was carried out by the flux method by melting under
  vacuum appropriately compacted mixtures of V$_2$O$_5$, V$_2$O$_3$ and NaVO$_3$
  in platinum crucibles and subsequently slow cooling of these melts
  from 1073 K to room temperature. Depending on the cooling parameters, 
  either needle-shaped or plated shaped crystals, up to 2 cm long,
  were obtained\cite{ueda}. 
  
 For checking purpose we refined also in space group {\em P2$_1$mn} which refinement
 converged to wR(F$^2$)=0.0512 for 1355 reflections with F$^2_0$$\geq$0 and
 50 parameters and R(F)=0.0201 for 1245 reflections obeying F$_0$$\geq$4.0 $\sigma$(F$_0$). 
 Inspection of the refined coordinates revealed a pseudo
 inversion center with the largest deviation for O1  of 0.03 \AA\cite{lapage,spek}. 
 The result of this refinement supports our result of the centro symmetric
 space group {\em Pmmn}.

 The refinements are similar in the sense that V is coordinated by a square 
 pyramid of O with the apical oxygen at smaller distance to V 
 (1.6150(9) \AA) than the ones at the base (1.8259(6)-1.9867(9) \AA) {\em versus} 
 V1 1.622(4) and 1.854(6)-1.966(5), Carpy and Galy\cite{carpy}: 1.65(5), 
 1.89(5)-1.96(5) \AA\ and for V2 1.604(6), 1.800(6)-2.007(4)\AA, 
 Carpy and Galy\cite{carpy}: 1.53(5), 1.76(5)-1.98(5) \AA).

\noindent
\begin{table}[htb]
  \begin{tabular}{ll}
{\em Crystal data}                 &                                          \\ \hline
$\alpha^{\prime}$-NaV$_2$O$_5$     &        D$_m$ not measured         \\
M$_r$=204.87                       &         Mo K$\alpha$ radiation     \\
Orthorhombic                       &         $\lambda$=0.71073 \AA\              \\
Pmmn                               &          Cell param. from 22 refl.                          \\
a=11.311(1) \AA\                   &          $\theta$=14.57-22.37$^{\circ}$         \\
b=3.610(1) \AA\                    &          $\mu$=4.77 mm$^{-1}$                 \\
c=4.800(1) \AA\                    &          T=295 K                            \\
V=196.00(7) \AA$^3$                &          0.20x0.15x0.013 mm$^3$            \\
Z=2                                &           Black                               \\
D$_x$=3.471 Mg m$^{-3}$            &            Crystal source: synthesis                      \\ \hline \hline
{\em Data collection}  &     \\ \hline
Enraf-Nonius CAD-4F diffract.    &         R$_{int}$=0.0232                 \\
$\omega/2\theta$ scans                &      $\theta_{max}$=39.96$^{\circ}$       \\
Abs. corr.: gaussian by integr.       &      h=-20$\rightarrow$20                 \\
T$_{min}$=0.6010; T$_{max}$=0.9381   &      k=0$\rightarrow$6                  \\
1472 measured reflections             &      l= 0$\rightarrow$8                 \\
701 independent reflections           &      frequency: 180 min                 \\
650 reflections with $>$2$\sigma$(I)  &     intensity decay: 1.0\%              \\  \hline \hline
{\em Refinement}                     &                                                     \\ \hline
Refinement on F$^2$                  &     ($\Delta/\sigma$)$_{max}$=0.001                  \\
R(F)=0.0151                          &      $\Delta\rho_{max}$=0.675 e \AA$^{-3}$           \\
wR(F$^2$)=0.0386                     &      $\Delta\rho_{min}$=-0.429 e \AA$^{-3}$          \\
S=1.126                              &      Extinction corr.: SHELXL                   \\
701 reflections                      &      Extinction coeff.: 0.067(4)                \\
28 parameters                        &      Scatt. factors: Cryst. Tab.        \\
  \end{tabular}
\end{table}

\begin{figure}[htb]
\centerline{\epsfig{figure=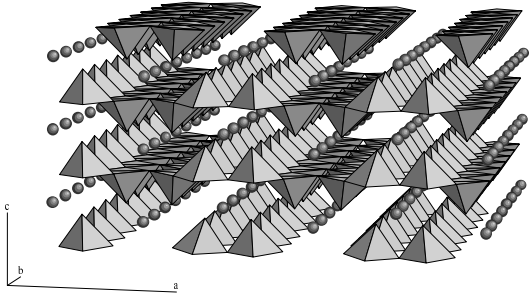,width=7.1cm,clip=}}
\caption{Crystal structure of $\alpha^{\prime}$-NaV$_2$O$_5$; square pyramids
around V (V-O1, V-O2, V-O3 x 3) and rows of Na.
}
\label{fig1}
\end{figure}

\noindent
\begin{figure}[htb]
\centerline{\epsfig{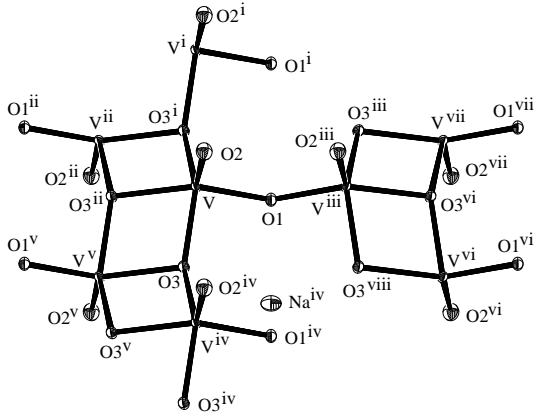}}
\caption{ Drawing of NaV$_2$O$_5$ showing several VO$_5$ square pyramids, with displacement 
ellipsoids at the 50\% probability level. [Symmetry codes:
             (i) $x,1\!+\!y,z$;                                                 
             (ii) $-x,\frac{1}{2}+y,1\!-\!z$;   
             (iii) $\frac{1}{2}\!-\!x,\frac{3}{2}\!-\!y,z$;                                           
             (iv) $x,-1\!+ y,z$; 
             (v)\,\,$-x,-\frac{1}{2}\!+\!y,1\!-\!z$;  
             (vi) $\frac{1}{2}\!+\!x,1\!-\!y,1\!-z$;                       
             (vii) $\frac{1}{2}\!+\!x,\frac{1}{2}\!+\!y,1\!+\!z$;
             (viii) $\frac{1}{2}\!-\!x,\frac{1}{2}\!-\!y,z$.]}
\label{fig2}
\end{figure}

\noindent
\begin{table}[htb]
  \begin{tabular}{lcccc}
   & $x$        &  $y$ & $z$         & U$_{eq}$      \\ 
V  & 0.09788(1) &  3/4 & 0.60781(30)  &  0.0073(1) \\
O1 & 1/4        &  3/4 & 0.4805 (20)   &  0.0094(2) \\
O2 & 0.11452(7) &  3/4 & 0.94197(17) &  0.0151(2) \\
O3 & 0.07302(6) &  1/4 & 0.48769(16) & 0.0097(1)  \\
Na & 1/4        &  1/4 & 0.14080(15) & 0.0170(2)  \\
  \end{tabular}
\medskip
\caption{Fractional atomic coordinates and equivalent isotropic parameters (\AA$^2$).}
\end{table}
\noindent

\noindent
\begin{table}[htb]
  \begin{tabular}{lrlr}
V-01     &    1.8259(6)    &     V-O3$^i$           &     1.9156(6)   \\
V-02     &    1.6150(9)    &     V-O3$^{ii}$        &     1.9867(9)   \\
V-03     &    1.9156(6)    &                        &                 \\ \hline
O1-V-02  &    102.86(4)    &     O3-V-O3$^{i}$      &     140.87(3)   \\
O1-V-03  &     92.15(2)    &     O3-V-O3$^{ii}$     &      77.74(2)   \\
O1-V-03$^{i}$  &    92.15(2)    &     O3$^{i}$-V-O3$^{ii}$      &     77.74(2)   \\
O1-V-03$^{ii}$  &    147.11(4)    &     V-O1-V$^{iii}$      &     140.89(6)   \\
O2-V-03  &    108.42(2)    &     V-O3-V$^{iv}$      &     140.87(4)   \\
O2-V-03$^{i}$  &    108.42(2)    &     V-O3-V$^{v}$      &     102.26(2)   \\
O2-V-03$^{ii}$  &    110.03(4)    &     V$^{iv}$-O3-V$^{v}$      &     102.26(2)   \\
  \end{tabular}
\medskip
\caption{Selected geometric parameters (\AA, $^{\circ}$).}
\end{table}

\end{document}